\documentclass[aps,prapplied,twocolumn,showpacs,amsmath,amssymb,superscriptaddress,floatfix,longbibliography,nofootinbib]{revtex4-2}

\usepackage{tikz}
\usepackage{graphicx}
\usepackage{dcolumn}
\usepackage{bm}
\usepackage{float}

\usepackage{svg}
\usepackage{xr}
\usepackage{braket}
\usepackage{bbold}
\usepackage{enumerate}
\usepackage{amsmath}
\usepackage{mathtools}
\usepackage{wrapfig}
\usepackage{amsmath} 
\usepackage{amssymb} 
\usepackage{seqsplit}
\usepackage{booktabs}
\usepackage{xcolor}
\usepackage{braket}
\usepackage{tabularx}
\usepackage{makecell}
\usepackage[export]{adjustbox}
\usepackage[shortlabels]{enumitem}
\usepackage{setspace}
\usepackage{tensor}
\usepackage{placeins}

\newcommand{\bes} {\begin{subequations}}
\newcommand{\ees} {\end{subequations}}
\newcommand{\beq}{\begin{equation}}
\newcommand{\eeq}{\end{equation}}

\newcommand{\tr}{{\rm Tr}}  
\newcommand*{\ketbra}[2]{\lvert #1 \rangle\!\langle #2 \rvert}
\def\theYear{\the\year}
\graphicspath{ {./figures/} }

\usepackage{hyperref}
\hypersetup{
    colorlinks=true,
    linkcolor=blue,
    citecolor=red,      
    urlcolor=cyan,
}
\usepackage[english]{babel}
\usepackage[capitalise]{cleveref}

\begin{document}

\title{Universal Hamiltonian control in a planar trimon circuit}
\author{Vivek Maurya}
\email{vmaurya@usc.edu}
\affiliation{Center for Quantum Information Science and Technology, University of Southern California, Los Angeles, California 90089}
\affiliation{Department of Physics \& Astronomy, University of Southern California, Los Angeles, California 90089}
\author{Daria Kowsari}
\affiliation{Center for Quantum Information Science and Technology, University of Southern California, Los Angeles, California 90089}
\affiliation{Department of Physics \& Astronomy, University of Southern California, Los Angeles, California 90089}

\author{Kumar Saurav}
\affiliation{Center for Quantum Information Science and Technology, University of Southern California, Los Angeles, California 90089}
\affiliation{Ming Hsieh Department of Electrical \& Computer Engineering, University of Southern California, Los Angeles, California 90089}

\author{S.A. Shanto}
\affiliation{Center for Quantum Information Science and Technology, University of Southern California, Los Angeles, California 90089}
\affiliation{Department of Physics \& Astronomy, University of Southern California, Los Angeles, California 90089}

\author{R. Vijay}
\affiliation{Department of Condensed Matter Physics and Materials Science,
Tata Institute of Fundamental Research, Homi Bhabha Road, Mumbai 400005, India}

\author{Daniel A. Lidar}
\affiliation{Center for Quantum Information Science and Technology, University of Southern California, Los Angeles, California 90089}
\affiliation{Department of Physics \& Astronomy, University of Southern California, Los Angeles, California 90089}
\affiliation{Ming Hsieh Department of Electrical \& Computer Engineering, University of Southern California, Los Angeles, California 90089}
\affiliation{Department of Chemistry, University of Southern California, Los Angeles, California 90089}
\affiliation{Quantum Elements, Inc, Westlake Village, California 91361}

\author{Eli M. Levenson-Falk}
\email{elevenso@usc.edu}
\affiliation{Center for Quantum Information Science and Technology, University of Southern California, Los Angeles, California 90089}
\affiliation{Department of Physics \& Astronomy, University of Southern California, Los Angeles, California 90089}
\affiliation{Ming Hsieh Department of Electrical \& Computer Engineering, University of Southern California, Los Angeles, California 90089}
\affiliation{Quantum Elements, Inc, Westlake Village, California 91361}

\begin{abstract}
    Multimode circuits provide an avenue for flexible control of single and multi-qubit gates. In this work we implement a multimode circuit known as a trimon integrated in a planar geometry. The trimon features three transmon-like modes with strong all-to-all $ZZ$ coupling. We demonstrate high fidelity operations on the trimon, achieving flexible control of its rich state space. This includes qubit rotations conditioned on one or both other qubits, unconditional single-qubit rotations, 
    and both excitation-conserving and double-excitation two-qubit entangling gates. Through multi-tone driving we are able to implement all 16 two-qubit Pauli operators in the two-qubit space. We further demonstrate using the trimon as a qudit with up to 8 states and higher coherence than typical transmon-based implementations. Our results show a compact, highly controllable device that can potentially replace transmons in standard superconducting processor architectures.

\end{abstract}

\maketitle
\section{Introduction}
Achieving precise and versatile multi-qubit control is 
essential for quantum simulation and computation, both of which 
necessitate
the ability to implement a wide range of Hamiltonian 
interactions---a requirement made all the more pressing in the NISQ era, where gate fidelities and circuit depths remain constrained.
In superconducting architectures, two-qubit gate schemes typically rely on auxiliary circuit elements such as tunable couplers~\cite{tunable_coupler_1,collodoImplementationConditionalPhase2020,tunable_coupler_4,tunable_coupler_5,tunable_coupler_6,kandalaDemonstrationHighFidelityCnot2021,tunable_coupler_9} and, 
in general, cannot directly implement all two-qubit Hamiltonian interactions.
While these engineered interactions can generate useful Hamiltonian terms, including exchange-type and $ZZ$ couplings, extending such control to a more complete set of two-qubit interactions remains a significant challenge.

Generally, always-on couplings 
result in unwanted terms 
that cause coherent errors.
Specifically in transmons~\cite{transmon-invention}, 
the weak anharmonicity causes 
couplings with higher energy levels that result in unwanted cross-Kerr interactions,
manifesting as static $ZZ$ coupling~\cite{staticZZ1,staticZZ2}. Such parasitic $ZZ$ terms often induce spectator errors and qubit dephasing, thereby degrading overall gate performance~\cite{ZZcoupling1,DDonZZ,ZZcoupling3,ZZcoupling4,Mundada}. While weak $ZZ$ can be mitigated with device architecture or via external drives~\cite{DDonZZ,kuSuppressionUnwanted2020,kandalaDemonstrationHighFidelityCnot2021,tripathi2021suppression}, these add complication and error if calibration is imprecise. 

On the other hand, \emph{strong} $ZZ$ coupling can be exploited as a useful resource to perform two-qubit gates. Moderately strong $ZZ$ interactions are typically used to implement controlled phase rotations~\cite{collodoImplementationConditionalPhase2020,liTunableCouplerRealizing2020,longUniversalQuantumGate2021,mitchellHardwareEfficientMicrowaveActivatedTunable2021,gossHighfidelityQutritEntangling2022,jinFastTunableHigh2024}. 
Even stronger $ZZ$ can separate conditional transition frequencies by an amount large compared to the achievable Rabi rate and pulse bandwidth, so that a drive resonant with one conditional transition is effectively off-resonant for the others.
Strong $ZZ$ thus allows a qubit's transition to be driven conditional on the other qubits' states. The resulting $n$-qubit gate is $C^{n-1}\mathcal{R}(\theta,\phi)$, a qubit rotation by angle $\theta$ about an axis in the $xy$ plane with azimuthal angle $\phi$, conditional on the $n-1$ other coupled qubits.
Such very strong dispersive couplings often arise in \emph{multimodal} superconducting circuits~\cite{PhysRevA.98.052318} (a.k.a. artificial molecules~\cite{kouFluxoniumBasedArtificialMolecule2017}), where modes of oscillation are hybridized by nonlinear couplings. A dispersive coupling between two modes will move one mode's transition frequencies depending on the number of excitations in the other mode. Treating each mode as a qubit, this behavior appears as a static $ZZ$ interaction that shifts each qubit’s transition frequency depending on the states of the other coupled qubits. 

In this work, we use a multi-mode circuit to demonstrate high-fidelity, flexible multi-qubit control. We implement a version of the three-mode \textit{trimon}~\cite{PhysRevA.98.052318} circuit integrated with two planar readout resonators and a charge drive line that enables fast control. This architecture natively supports three-qubit conditional rotation CC$\mathcal{R}(\theta,\phi)$ gates due to strong inter-mode dispersive couplings. We demonstrate that it is straightforward to implement complete control of the trimon, performing fast single- and two-qubit control using simultaneous pulses applied through the drive line. 
We 
also
demonstrate gates mediated by second order stimulated Raman transitions~\cite{raman4,Raman1,Raman3,Raman2}. 
We use these Raman transitions to perform high-fidelity $\sqrt{i\text{SWAP}}$ (excitation-conserving) and $\sqrt{i\text{bSWAP}}$~\cite{PhysRevLett.109.240505} (double-excitation) gates 
in a two-qubit subspace (with the third mode prepared in $\ket{0}$), with SPAM-corrected QPT gate fidelities of $99.19\%$ and $99.59\%$, respectively. 
We further demonstrate virtual controlled-phase gates that implement any diagonal unitary in the three-qubit space, including $\text{CZ}$ and $\text{CCZ}$, implemented via independent updates of the rotating-frame phases associated with the conditional transitions.
Together, these results 
provide evidence 
that the device is capable of realizing a broad class of two-qubit Hamiltonians via a \emph{single} 
shaped envelope (possibly containing multiple simultaneous frequency components). The control can implement Hamiltonian terms that span the $15$-dimensional traceless two-qubit operator space (equivalently, the $15$ non-identity two-qubit Pauli products).
Beyond multi-qubit gates, we demonstrate the versatility of this platform by using it as a qudit. We show high-fidelity control by implementing single axis dynamical decoupling sequences on qutrit, ququart, six-level (qud6), and eight-level (qud8) subspaces. Our results show that the trimon can be used for high-fidelity multi-qubit operations and could potentially replace transmons in quantum processors. We 
conclude with a 
discussion of 
scalability of the trimon and its potential use in quantum processors. 

\section{Device Description}The trimon consists of a ring of four Josephson junctions with capacitance between all circuit nodes and to ground. The simplified Hamiltonian for a trimon can be expressed as
\begin{equation}
\mathcal{H}=\sum_{\mu=A, B, C}\left[\omega_\mu \hat{n}_\mu-J_\mu \hat{n}_\mu^2\right]-2 \sum_{\mu>\nu} J_{\mu \nu} \hat{n}_\mu \hat{n}_\nu
\label{Hamiltonian}
\end{equation} 
where $\hat{n}_\mu$ is the excitation number operator, $\omega_\mu$ is a mode frequency parameter (so that in this convention the $\ket{0}\!\rightarrow\!\ket{1}$ transition frequency of mode $\mu$, with the other modes unexcited, is $\omega_\mu-J_\mu$), $-2J_\mu$ is its anharmonicity, and $-2J_{\mu\nu}$ is the cross-Kerr (dispersive) shift per excitation between qubit modes $\mu$ and $\nu$ (see Appendix \ref{device_theory} for a derivation of this effective Hamiltonian). 
Hereafter, we refer to the modes as separate qubits, although they are transmon-like and thus have higher levels. 
Because the interaction is diagonal in the excitation-number basis (i.e., it commutes with the uncoupled number operators), it produces purely longitudinal (dispersive) energy shifts and does not induce excitation exchange or hybridize the eigenstates in this simplified model.
This corresponds to strong effective $ZZ$-type interactions in the qubit subspace, i.e., conditional transition-frequency splittings $2J_{\mu\nu}$ on the order of a few hundred MHz.
Consequently, each qubit exhibits four distinct transition frequencies corresponding to the different states of the other two qubits. We denote these frequencies by a subscript with numbers for the undriven qubit states and a letter for the driven mode, e.g., $\omega_{1\mathrm{B}0}$ is the transition frequency of qubit B with qubit A in the $\ket{1}$ state and qubit C in the $\ket{0}$ state. The values of all allowed dipolar transition frequencies are provided in \cref{fig:trimon_circuit}.
For the Rabi rates and pulse bandwidths used here, these conditional splittings are large enough that a tone resonant with one conditional transition produces negligible population transfer on the other conditional transitions. Residual off-resonant effects (primarily AC Stark shifts) are compensated using virtual phase updates.
A charge drive thus implements a 3-qubit conditional rotation, e.g., $\ket{100} \leftrightarrow \ket{110}$ for $\omega_{1\mathrm{B}0}$. 
This is in contrast to more conventional superconducting qubit architectures, where each qubit has (to a good approximation) a single addressable transition frequency and single-qubit rotations $\mathcal{R}(\theta,\phi)$ are directly implemented; here, strong cross-Kerr shifts split each mode into multiple conditionally addressable transitions. A reference clock in the control unit is used to track the frequency of each addressable transition. Importantly, the required number of reference clocks scales only linearly with the number of trimon modes and can be accommodated by typical commercially available FPGAs.
A single-qubit rotation can be realized using four native CC$\mathcal{R(\theta,\phi)}$ gates with all four control conditions (or two control conditions for a two-qubit rotation). 
In the ideal block-diagonal limit, these CC$\mathcal{R}$ operations commute and may be applied in series or simultaneously; in practice, simultaneous driving induces AC Stark shifts of the conditional transitions, which we compensate using calibrated detunings and virtual phase updates.

\begin{figure}[h!]
\centering
\includegraphics[width=1\columnwidth]{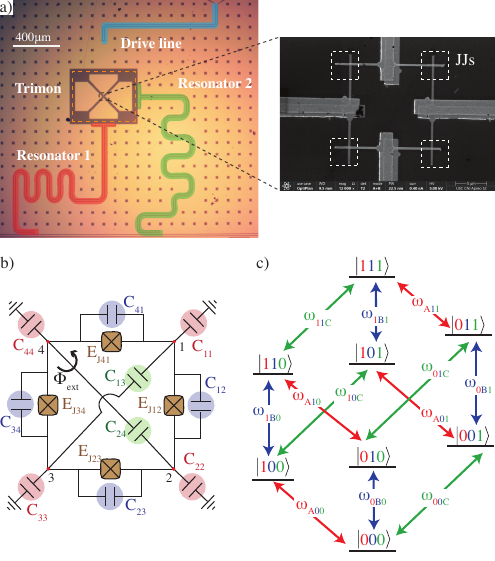}
\caption{(a) Optical image of the planar trimon device. The central trimon circuit (orange box) consists of four capacitor pads connected via four Josephson junctions. (Right) A zoomed-in Scanning Electron Microscope (SEM) image of the junctions at the center of the circuit. Two $\lambda/4$ coplanar waveguide readout resonators (red and green) are capacitively coupled to adjacent trimon pads, while a dedicated drive line (blue) couples capacitively to the circuit to enable direct control of the modes. (b) Lumped-element circuit model of the planar trimon. Self capacitances (red) represent the shunt capacitance from each pad to ground. Adjacent pad capacitances (purple) arise from nearest neighbor coupling between the pads. Cross-capacitances (green) correspond to coupling between opposite pads and are approximately an order of magnitude smaller than the self capacitance. The brown elements denote the Josephson junctions connecting the four pads. Numbers 1, 2, 3, and 4 (red dots) indicate the four nodes of the circuit. Capacitance and Josephson energy values are listed in Appendix \ref{app:design}. (c) Energy level diagram of the three-qubit subspace, showing the $12$ allowed dipole transitions between the $8$ computational basis states. Transitions associated with the A, B, and C modes are indicated in red, blue, and green, respectively. The values of these frequencies are provided in \cref{fig:trimon_circuit}.}
\label{fig1: circuit}
\end{figure}

An optical image and a circuit schematic of our planar trimon device is given in \cref{fig1: circuit}. The trimon consists of four capacitor pads embedded in a large ground plane. Neighboring pads are connected via a ring of four Josephson junctions, all with approximately equal Josephson energy. 
Opposite capacitor pads are nominally identical, ensuring minimal geometric asymmetry; any residual asymmetry arises from 
fabrication disorder (e.g., junction-to-junction variation) and from coupling to the drive line and readout resonators.
These are $\lambda/4$ meandering coplanar waveguide (CPW) resonators, each capacitively coupled to adjacent pads of the trimon. A charge line embedded in the ground plane provides capacitive coupling to the entire circuit, enabling direct driving of the trimon modes. Assuming complete symmetry, the device modes are ``dipolar" charge oscillations between opposite capacitor pads (i.e., nodes 1 and 3 for one mode, nodes 2 and 4 for another) and a ``quadrupolar" oscillation between neighboring nodes (i.e. nodes 1 and 3 together versus 2 and 4 together) \cite{PhysRevA.98.052318}. Device parameters are given in \cref{table: device-parameters} and more details of the device design are given in Appendix \ref{app:design}.

A similar device has been previously demonstrated in a 3D cavity architecture setup~\cite{PhysRevApplied.7.054025, PhysRevApplied.14.014072}. 
However, 
the lack of a dedicated drive channel strongly coupled to all modes, together with the weak coupling of some trimon modes to the 3D cavity field,
leads to comparatively long gate times. 
In a single 3D cavity, the non-aligned dipolar and quadrupolar modes in the trimon couple even more weakly to the cavity field than the aligned dipolar mode, greatly complicating qubit control and readout. In contrast, in the planar-embedded device, a charge drive line and cavity mode(s) strongly coupled to all trimon modes are trivial to implement. As we show below, the strong driving allows us to pursue much more flexible control schemes. The planar device is also much more compact due to the large capacitance to ground from each pad. 

\section{Readout}
We use ordinary circuit QED dispersive readout for the trimon, similar to standard transmon devices. In the particular device reported here, states with the same total excitation number yielded nearly identical responses on both readout resonators. The resonator responses thus do not uniquely distinguish all $8$ computational states in a single shot. This was a deliberate design choice, as the device was originally designed to explore other physics requiring overlapping dispersive shifts. In the current work, readout of all states
then
requires four measurement rounds with mapping pulses ~\cite{PhysRevA.98.052318, PhysRevApplied.7.054025}. The mapping pulses map a pair of states with overlapping readout responses to a pair of well-separated readout states. 
Our control electronics 
further limit the number of distinct pulses that can be defined in a single experiment, preventing readout of all three qubits within the same experimental sequence. We therefore restrict readout to two qubits at a time. 

This overlap of dispersive shifts occurs even though each resonator nominally only couples to 2 of the modes (resonator 1 to A and C, 2 to B and C). This is because of the inter-mode $ZZ$ interactions. For instance, changing mode A's state changes the energies of modes B and C, altering their detunings from resonator 2 and thus their Lamb shifts. In this device the net effect of those Lamb shift changes is identical to the dispersive shift from a mode that is coupled to the resonator. This degeneracy is particular to the exact parameters of this device, and a different choice of resonator frequencies and couplings would lift this degeneracy. In principle, a device optimized for three-qubit readout with sufficiently non-degenerate dispersive responses across the two resonators should be able to distinguish all $8$ computational states in a single shot using the two complex resonator responses (four real quadratures); see Appendix \ref{app:design}.
These limitations are not intrinsic to the trimon and can be avoided in future experiments by optimizing the readout design (to yield nondegenerate dispersive responses) and by using more flexible control electronics; for this study, we restrict to single- and two-qubit readout. More details of the readout procedure, including the mapping pulses used and the analysis procedure, are given in Appendix \ref{readout-appendix}.
We next discuss our results.
\begin{figure}[h!]
\centering
\includegraphics[width=1.0\columnwidth]{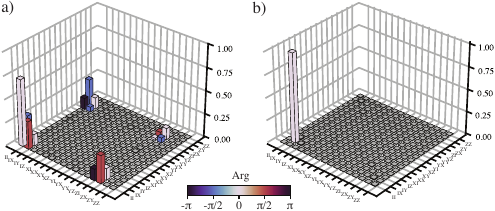}
\caption{SPAM corrected two-qubit QPT results for a $60$ns (a) $cX_{\pi/2}$  which implements a $X_{\pi/2}$ rotation on qubit B conditioned on qubit A being in $\ket{0}$, and (b) $X_{\pi}$, realized by applying simultaneous drives at $\omega_{0B0}$ and $\omega_{1B0}$. Qubit C is kept in the ground state throughout. When embedded in the full three-qubit computational subspace (with qubit C prepared in the ground state), these gates correspond to $ccX_{\pi/2}$ and a $cX_\pi$ gate, respectively. Bar heights indicate matrix element magnitude, and color indicates the associated complex phase; the color of amplitudes smaller than $0.01$ 
is
suppressed for clarity.}
\label{fig:QPT_CNOT}
\end{figure}

\section{Experimental Results}
\subsection{Multi-Qubit Gates}
The 12 trimon dipole transitions each have their own frequencies and thus can each have their own phase references, Bloch spheres, and pulse definitions. We drive a transition to achieve a three-qubit conditional rotation. This causes off-resonant driving of all other transitions, leading to AC Stark shifts which induce 
deterministic diagonal phase shifts on computational states. We compensate these phase shifts using software-defined phase (``virtual $Z$'') updates. Due to the readout 
limitations
described above, we keep qubit C in its ground state and focus only on the 2-qubit space (A and B). 
\Cref{fig:QPT_CNOT} shows two-qubit (A, B) quantum process tomography (QPT), with qubit C prepared in $\ket{0}$, of a $ccX_{\pi/2}^{0B0}$ gate \cite{EasybetterQPT}. 
The gate fidelity, computed from QPT and corrected for state preparation and measurement (SPAM) error, is $99.87\%$ (see Appendix \ref{section-tomo} for the SPAM correction procedure). QPT can detect any errors arising from unwanted conditionality (or lack thereof) in the gate, so this high fidelity indicates that our gate is indeed implementing a near-ideal CCNOT.
We also perform randomized benchmarking (RB) and interleaved RB 
on the conditional $\omega_{0B0}$ transition (i.e., within the corresponding two-level subspace),
obtaining an average RB gate fidelity of $99.93\%$ for a $40$ns CC$\mathcal{R}$ operation. RB gate fidelities for $\omega_{A00}$ and $\omega_{00C}$ along with calibration methods of other dipolar transitions are given in Appendix \ref{Gate-calibration}. 

While conditional gates require only a single pulse and exhibit high fidelity, implementing unconditional operations requires combining multiple pulses with opposite control conditions. When applied sequentially, this approach doubles or quadruples the gate duration, increasing incoherent errors from decay and dephasing. Instead, we realize two-qubit C$\mathcal{R}$ operations by applying two CC$\mathcal{R}$ pulses simultaneously with opposite conditions on the uninvolved qubit. 
Because the two CC$\mathcal{R}$ drives address orthogonal control subspaces, the corresponding ideal operations commute; after compensating AC Stark shifts, simultaneous application removes the conditionality on the uninvolved qubit without increasing the gate duration.
Again restricting readout to qubits A and B, \cref{fig:QPT_CNOT}b shows the SPAM-corrected QPT of a $cX_\pi^{B0}$ gate on qubit B conditioned on qubit C being in $\ket{0}$ but unconditional on qubit A (i.e., acting as an unconditional $X_\pi$ on B within the $C=\ket{0}$ manifold), implemented by applying $ccX_\pi^{0B0}$ and $ccX_\pi^{1B0}$ simultaneously. The resulting QPT gate fidelity is $99.76\%$. 
Each drive induces AC Stark shifts of the other transition frequency, requiring additional drive detuning, which we correct for in our deterministic benchmarking (DB) gate calibration~\cite{DB} (see Appendix \ref{Gate-calibration}). Again, QPT shows no sign of conditionality on mode A, indicating that the $ZZ$ interaction is not causing error and that this is a proper gate that requires no knowledge of the mode A state.

Similarly, a fully unconditional $\mathcal{R}$ operation on a qubit can be realized using four simultaneous CC$\mathcal{R}$ gates with appropriate AC Stark shift corrections. We calibrated such an operation with the same duration as the original CC$\mathcal{R}$ gate. To characterize its performance, we perform a single-qubit QPT on qubit B. The system is initialized in the state $\ket{+}_A \ket{0}_B \ket{+}_C$, preparing qubits A and C in equal superpositions while keeping qubit B in the ground state. For this experiment, we use a different readout procedure that measures only qubit B and traces over the states of qubits A and C (see Appendix \ref{readout-appendix}), thereby isolating the effective action of the unconditional operation on the target qubit. Using this approach, we obtain a SPAM-corrected QPT gate fidelity of $99.56\%$ (see Appendix \ref{Gate-calibration}). This characterization does not include any accidental effects on qubit A and C transitions. We expect these to take the form of AC Stark shifts that could be corrected with calibration, as we do in our CC$\mathcal{R}$ and C$\mathcal{R}$ gates. A more complete characterization of the fully unconditional gate would require simultaneous readout of all three qubits, which we leave for future work.  

Because each conditional rotation is driven with a separate tone, we can change the phase of each individually. This changes the axis of rotation for each $ccR$ gate. 
We can thus implement conditional or unconditional $Z$ rotation gates virtually by shifting these drive phases~\cite{vezvaee2024virtualzgatessymmetric}. 
More generally, because we independently track the phases of the conditional transitions, these phase updates implement arbitrary diagonal unitaries on the computational basis, including $\text{CZ}$ and $\text{CCZ}$.
As software-defined frame updates, these operations are expected to be ideal; experimentally, SPAM-corrected two-qubit QPT is consistent with the ideal intended operation to within our measurement precision.
We use these virtual $Z$ rotations after driving a CC$R$ gate to correct AC Stark shifts that occur on the other $11$ allowed transitions within the computational manifold, and when driving a C$R$ gate to correct shifts on the other $10$ transitions [recall that there are $12$ allowed dipole transitions and CC$R$ (C$R$) targets one (two) of those transitions]. We also use them to counteract unwanted relative phase differences that occur due to the different propagation times for different frequencies to travel from our electronics to the device; more details are given in Appendix \ref{Gate-calibration}.

\subsection{Generating Entangled States}
\begin{figure}[t]
\centering
\includegraphics[width=1\columnwidth]{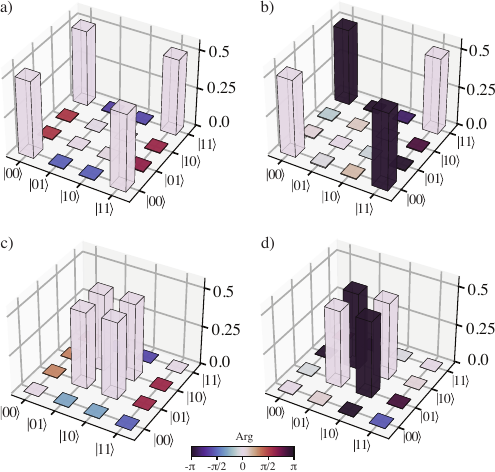}
\caption{Tomographic reconstruction of the four Bell states: (a) $\ket{\Phi_{+}}$, (b) $\ket{\Phi_{-}}$, (c) $\ket{\Psi_{+}}$, and (d) $\ket{\Psi_{-}}$, obtained after applying REM and MLE (see Appendix \ref{section-tomo}). The bars indicate the magnitude of each matrix element, while the color denotes its phase. }
\label{fig:QST}
\end{figure}

Native CC$\mathcal{R}$ rotations make the generation of entangled states straightforward. The maximally entangled Bell states can be prepared using only two pulses (and GHZ states or W states can be prepared using three pulses, although we do not demonstrate that here). \Cref{fig:QST} shows two-qubit state tomography of all four Bell states $\ket{\Phi_\pm}\equiv (\ket{00}\pm\ket{11})/\sqrt{2}$, $\ket{\Psi_\pm}\equiv (\ket{01}\pm\ket{10})/\sqrt{2}$, corrected for SPAM errors (see Appendix \ref{section-tomo}). Each Bell state is prepared using a sequence consisting of a CC$\mathcal{R}(\pi/2,\phi)$ gate followed by a CC$\mathcal{R}(\pi,\phi)$ gate, where the conditionality and the phase $\phi$ are chosen according to the target Bell state. The resulting state fidelities are $99.67\%$ for $\ket{\Phi_{+}}$, $99.93\%$ for $\ket{\Phi_{-}}$, $99.94\%$ for $\ket{\Psi_{+}}$, and $99.38\%$ for $\ket{\Psi_{-}}$, all with concurrence greater than $0.99$. Note that the different fidelities of $\ket{\Phi_+}$ vs $\ket{\Phi_-}$ and $\ket{\Psi_+}$ vs $\ket{\Psi_-}$ merely reflect small calibration drifts between taking data, as the definitions of these states are arbitrary. These fidelities are SPAM-corrected using 
readout error mitigation (REM) and maximum likelihood estimation (MLE) 
and thus depend on the mitigation/reconstruction model, but the high gate-fidelity demonstrated via QPT and RB 
(which we report as independent validation) 
further confirms that we can generate nearly pure entanglement with simple pulses. Moreover, we 
next 
demonstrate the capability to generate a two-qubit maximally entangled state (a Bell state up to a relative phase) with just one pulse.

\subsection{Raman Gates}
The strong driving possible with a planar trimon enables higher-order processes such as two-photon Raman transitions. We drive these processes by driving transitions between two ``active'' states and a common intermediate state, 
with equal single-photon detuning $\Delta$. The drives satisfy the two-photon resonance condition: the difference between the two drive frequencies (or their sum for pair processes) matches the active-state splitting. In the regime where detuning is large compared to the two drive strengths and the intermediate level's linewidth $|\Delta|\gg \Omega_1,\Omega_2, \gamma$, 
the intermediate state is only virtually populated. In this case the dynamics reduce to an effective two-level Hamiltonian between the active states with coupling $\Omega_{\mathrm{eff}}\sim \Omega_1\Omega_2/(2\Delta)$, accompanied by AC Stark shifts $\propto \Omega_{1,2}^2/\Delta$ that we compensate with virtual phase updates.

\begin{figure}[t]
\centering
\includegraphics[width=1.0\columnwidth]{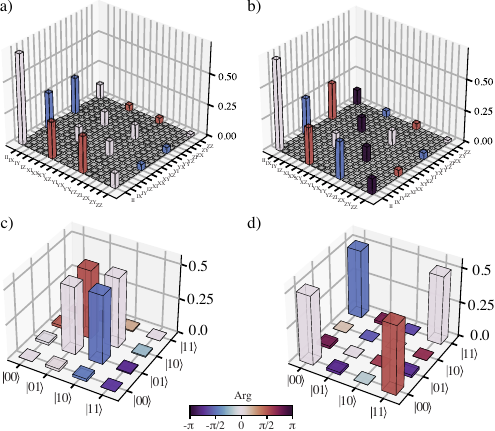}
\caption{Quantum process and state tomography of Raman-mediated entangling gates. (a) Process matrix of the $\sqrt{i\text{SWAP}}$ gate and (b) of the $\sqrt{i\text{bSWAP}}$ gate. (c) Reconstructed density matrix of the state $(\ket{01}-i\ket{10})/\sqrt{2}$ prepared by initializing $\ket{100}$ and applying the $\sqrt{i\text{SWAP}}$ gate. (d) Reconstructed density matrix of the state $(\ket{00}+i\ket{11})/\sqrt{2}$ produced directly using a single $\sqrt{i\text{bSWAP}}$ gate. Bar height indicates the magnitude of each matrix element and color denotes its phase; for QPT, the color of amplitudes smaller than $0.015$ is suppressed for clarity. }
\label{fig:QPT_raman}
\end{figure}

Restricting ourselves to two qubits A and B and keeping qubit C in $\ket{0}$, we realize an excitation-conserving $\sqrt{i\text{SWAP}}$ gate by simultaneously driving $\omega_{A00}$ and $\omega_{0B0}$ with a detuning $\Delta=32$ MHz. This is a rotation in the $\ket{10}\leftrightarrow\ket{01}$ space using $\ket{00}$ as the virtual intermediate state. Likewise, to realize a double-excitation $\sqrt{i\text{bSWAP}}$ gate, we drive $\omega_{0B0}$ and $\omega_{A10}$ with a detuning $\Delta=30$ MHz, a rotation in the $\ket{00}\leftrightarrow\ket{11}$ space using $\ket{01}$ as the virtual intermediate state. The resulting SPAM-corrected QPT $\chi$ matrices for both processes are shown in \cref{fig:QPT_raman}. We achieve SPAM-corrected gate fidelities of $99.19\%$ and $99.59\%$ 
for these two processes, respectively. 
We also perform state tomography of the states obtained after applying the $\sqrt{i\text{SWAP}}$ and $\sqrt{i\text{bSWAP}}$ gate to $\ket{10}$ and $\ket{00}$, respectively (with qubit C prepared in $\ket{0}$). The state fidelities of $(\ket{01}-i\ket{10})/\sqrt{2}$ and $(\ket{00}+i\ket{11})/\sqrt{2}$, after REM and MLE, were $99.87\%$ and $99.93\%$, respectively (see Appendix \ref{section-tomo}). 
Furthermore, the corresponding reconstructed density matrices show that the intermediate state population 
is below $0.1\%$ at the end of the Raman pulse for both processes.
With the high fidelity $\sqrt{i\text{bSWAP}}$ gate, we have the ability to generate a maximally entangled state $\ket{\Phi_+}$ with \emph{just one pulse}. Both of these Raman gates use $120$ ns cosine DRAG pulse profiles. Gate durations could be further reduced on this device by adopting flat-top profiles or increasing drive amplitude, and fidelities could be further improved using more sophisticated Raman control techniques (see Discussion).

Using a single Raman pulse at the $\sqrt{i\mathrm{SWAP}}$ gate frequency plus a virtual CPhase rotation, we can implement a continuous fermionic simulation (fSim) gate (a continuous $\ket{01}\leftrightarrow\ket{10}$ rotation combined with a continuous phase shift on $\ket{11}$)~\cite{googleaiquantumDemonstratingContinuousSet2020} natively on our architecture. 
Moreover, by combining exchange-type and pair-type Raman interactions (with appropriate amplitudes and phases), we can implement the $B$ gate $\exp[i\frac{\pi}{8}(2XX-YY)]$, which produces an arbitrary two-qubit unitary with the minimum number of discrete two-qubit gates ~\cite{Zhang:03}. We have demonstrated these gates but have not optimized them for gate time, Raman detuning $\Delta$, or fidelity;
a rapid calibration procedure for fSim and B gates will be the subject of future work. We emphasize that these gates can be implemented as a single operation and do not require concatenating multiple pulses.

\subsection{Arbitrary Hamiltonian Control}
With the drives used for our gates, we can realize 
an effective Hamiltonian that 
spans the two-qubit operator space. The $16$ Pauli products $\{I,X,Y,Z\}\otimes\{I,X,Y,Z\}$ form an operator basis. A single conditional drive realizes a projector-controlled rotation, e.g., $\ket{0}\!\!\bra{0}_A\otimes(\cos\phi\,X_B+\sin\phi\,Y_B)=\tfrac12(I+Z)_A\otimes(\cos\phi\,X_B+\sin\phi\,Y_B)$,
and similarly for $\ket{1}\!\!\bra{1}_A=\tfrac12(I-Z)_A$. By combining the two conditional tones in a single shaped pulse (with programmable relative phase and amplitude), we can isolate either unconditional $I\otimes(X/Y)$ terms or two-qubit $Z\otimes(X/Y)$ terms (and likewise exchanging $A$ and $B$). Idling implements $II$, and virtual frame updates provide $ZI$, $IZ$, and $ZZ$. In addition, Raman drives generate exchange and pair couplings, yielding Hamiltonians proportional to $XX+YY$ and $XY-YX$ ($i$SWAP Raman) and to $XX-YY$ and $XY+YX$ ($i$bSWAP Raman). Taken together, these controls span the space of two-qubit matrices, allowing an arbitrary effective two-qubit Hamiltonian to be realized using a single shaped envelope comprising one to a few frequency components applied through the common drive line.


\subsection{Trimon as a Qudit}
\begin{figure}[h!]
\centering
\includegraphics[width=1\columnwidth]{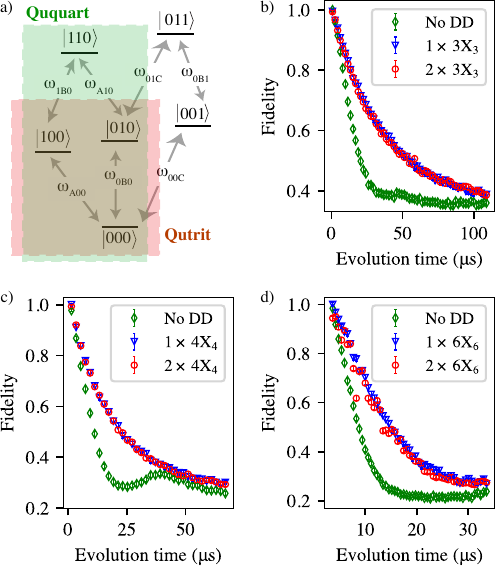}
\caption{(a) Energy level structure used in the qudit dynamical decoupling (DD) experiments. The red shaded box indicates the three levels forming the qutrit, the green-shaded box indicates the four-level ququart, and all six levels together constitute the qud6 system. (b-d) Experimental results for the qutrit, ququart, and qud6 single-axis DD sequence. The three sequences shown in each plot were interleaved and executed as a single experiment. For each plot, the fidelity is normalized by the maximum fidelity achieved in that experiment, defined as the maximum value among the three sequences. In each case, an effective $X$ pulse on the qudit space is decomposed using $2$, $3$ and $5$ native CC$X$ gates, corresponding to the qutrit, ququart, and qud6, respectively. Error bars correspond to $\pm2\sigma$.}
\label{fig:qudit_DD}
\end{figure}

Instead of viewing the trimon as a three-qubit system with strong all-to-all $ZZ$ coupling, we can also think of it as 
an effective qudit within the eight-state computational manifold spanned by $\{\ket{abc}\,|\,a,b,c\in\{0,1\}\}$.
Qudit systems implemented using higher excited states of standard transmons are generally more susceptible to decoherence due to enhancement of bath coupling matrix elements~\cite{PhysRevLett.134.050601}. Moreover, to compensate for the shallower potential associated with using these higher levels, transmon qudits are typically designed with relatively large $E_J/E_C$ ratios, which in turn reduces the anharmonicity and constrains the achievable gate speeds~\cite{wangHighE_JE_CTransmon2025}. In contrast, a trimon is expected to exhibit reduced sensitivity to such noise sources 
because the qudit manifold is constructed from the lowest two levels of each mode, avoiding population of higher excited states.
To demonstrate qudit control, we select an arbitrary subspace of dimension $d\leq 8$ and prepare an equal superposition of all $d$ qudit states. Using the approach outlined in Ref.~\cite{PhysRevLett.134.050601}, we then drive a qudit dynamical decoupling (DD) sequence to suppress quasi-static noise. We create a generalized $X_d$ gate which maps $\ket{0}\rightarrow\ket{1}\rightarrow\cdots\rightarrow\ket{d-1}\rightarrow\ket{0}$
using $d-1$ pulses; 
each DD cycle consists of $d$ repetitions of this gate (so that $(X_d)^d=I$). We then define a DD sequence $dX_d \equiv X_d f_\tau X_d f_\tau \cdots X_d f_\tau$, where $f_\tau$ is free evolution for a time $\tau$. We repeat this cycle $n$ times to form an $n\times dX_d$ sequence (see \cref{table:dd} for gate and sequence definitions).
\Cref{fig:qudit_DD} shows the results for $d=3, 4$, and $6$ levels; results for $d=8$ and details of the gates are in Appendix \ref{Qudit-DD-appendix}.  As expected, the applied DD sequences suppress coherent oscillations arising from frequency detuning, resulting in purely exponential decay. 
The decay time is also significantly improved relative to free evolution, in agreement with earlier DD work using transmon qubits~\cite{Pokharel2018,DD-survey,tripathi2021suppression}.
We 
find that increasing the number of repetitions $n$ in the $n \times dX_d$ sequence does not improve or degrade the decay time. 
This behavior is consistent with a noise spectrum dominated by very low-frequency components that are largely averaged by a single $dX_d$ cycle~\cite{ShiokawaLidar:02}, and/or by higher-frequency components beyond the effective bandwidth of the DD filter function~\cite{Kofman:01,Gordon:2008:010403}; in our device, flux noise is a likely contributor to the low-frequency dephasing. Given this noise spectrum, we would expect additional sequence repetitions to degrade fidelity if there were any significant coherent errors in the gates. We therefore conclude that our qudit gates are well calibrated.
We note that the coherence time of a $d=8$ equal superposition state is only $\sim2\times$ shorter than each of the single-qubit coherences, in contrast to transmon-based qudits, where decoherence rates typically scale approximately linearly with increasing level number~\cite{wangHighE_JE_CTransmon2025}.

\section{Discussion}
We have demonstrated a compact, planar, three-mode superconducting circuit where the modes can be operated as three transmon-like qubits or as a single collective qudit. The strong cross-Kerr couplings between modes allow for native three-qubit conditional rotation gates with fidelities exceeding 99.9\%, with performance primarily limited by decoherence for a 40 ns gate. By combining simultaneous drives, we can selectively remove the conditionality, implementing unconditional two-qubit and single-qubit rotations with minimal loss of fidelity. 
Moreover, two-tone Raman drives allow us to achieve 
$i$SWAP- and $i$bSWAP-like rotations
in a two-qubit subspace (with qubit C prepared in $\ket{0}$): we demonstrate a $\sqrt{i\mathrm{SWAP}}$ and a $\sqrt{i\mathrm{bSWAP}}$ gate with 99.19\% and 99.59\% fidelity, respectively, with simple pulse shapes, which can be improved using stimulated Raman adiabatic passage (STIRAP) or optimal control techniques~\cite{ramanoptimalcontrol}.

With conditional, unconditional, and Raman gates, combined with virtual $Z$ and $CZ$ rotations, we demonstrate 
the ability to 
realize effective Hamiltonians spanning the two-qubit operator space, enabling complete control of an effective two-qubit Hamiltonian using a single shaped envelope comprising one to a few simultaneous frequency components.
Such an expanded gate set has multiple potential applications, for example, in quantum sensing protocols requiring continuous control of entangled states~\cite{hechtBeatingRamseyLimit2025} and reduced circuit depth in quantum simulation algorithms.  
Further calibrating additional two-tone Raman transitions should expand the set of directly accessible three-qubit interactions and may enable 
realization of a broad class 
of effective three-qubit Hamiltonians
using a small number of simultaneous frequency components within a single shaped envelope.
The two-qubit Raman gates we demonstrate were all calibrated with qubit C in its ground state, and were conditional on that state. 
We therefore actually demonstrated three-qubit gates $\mathrm{C}\sqrt{i\mathrm{SWAP}}$ 
and $\mathrm{C}\sqrt{i\mathrm{bSWAP}}$ (conditional on qubit C); $\mathrm{C}\sqrt{i\mathrm{SWAP}}$ is a controlled partial-exchange operation related to the controlled-SWAP (Fredkin) gate~\cite{fredkinConservativeLogic1982}, which can improve algorithmic performance~\cite{parkReducingTCountQuantum2024}.
Our results demonstrate that high-fidelity, flexible multi-qubit and qudit control is possible in a compact circuit, and that the trimon may be a viable alternative platform for superconducting quantum processors.

The multimode structure of trimons makes them promising platforms for qudit-based Hamiltonian simulation. Qutrits have been proposed as natural simulation units for spin-1 systems~\cite{ogunkoya2024qutrit}, while higher-dimensional qudits have recently been employed in simulations of lattice gauge theories~\cite{qudits-lgt,john2026nonabelianstringbreakingdynamicsqudit}, lattice quantum chromodynamics, and fermionic systems~\cite{qu8its-lqc}, among other applications.

The control pulses used in this device were intentionally chosen to operate on the decoherence-limited side, prioritizing simplicity over gate speed. Even in this regime, the achieved gate speeds were substantially higher than 3D cavity-based trimon devices~\cite{PhysRevApplied.14.014072, PhysRevA.98.052318}. This choice was made due to the relatively low anharmonicity of the present device ($\sim$130 MHz), which limited how short we could make gate times with simple Gaussian DRAG pulses before leaking to other levels of the trimon. Pushing to faster gates would require more sophisticated optimal-control techniques to suppress leakage, which we leave to future work; higher-anharmonicity devices, comparable to typical transmons, could likewise relax this constraint. Optimal control in trimons with higher anharmonicity will be the subject of future work.

Aside from the high fidelity control we demonstrate in this work, using trimons may reduce wiring 
complexity by reducing the number of dedicated microwave control channels: a single charge drive line naturally can address three trimon modes. Similarly multiplexing the charge drive of three transmons typically would require careful engineering.  
Likewise, readout hardware can also be reduced in some designs (e.g., by jointly discriminating multi-mode states using two resonators), though the practical wiring advantage depends on the chosen multiplexing and readout architecture~\cite{mohseniHowBuildQuantum2025}.

Given that there is a large overlap between the design, fabrication, driving, and coupling schemes used for trimons and transmons, it would be straightforward to integrate trimons into current superconducting processor architectures to either replace or augment transmons. The design and fabrication of the current device can be further optimized, and there are clear paths to improving noise characteristics of the trimon. Firstly, the relaxation times were limited by a drive line coupling too strongly to mode B and capacitor pads coupling too much electric field into lossy interface layers. An optimized design should achieve coherence similar to a transmon fabricated with the same process, as the two devices have very similar matrix elements~\cite{PhysRevA.98.052318}. Consequently, any improvement in fabrication and materials for transmons will translate to improvements for trimons. Secondly, the trimon is flux sensitive due to a ring of junctions, which enables tunability of transition frequencies but is also a primary contributor to low-frequency dephasing in our system. To characterize this, we performed CPMG-N sequences with varying N and observed an improvement from Ramsey to CPMG-1 (Hahn echo), beyond which the coherence times saturated, consistent with our qudit experiments. This trend was seen across different flux points, with Ramsey times showing a lower baseline but CPMG decay times remaining comparable. Notably, this flux tuning is not necessary, and in fact we operate at the integer flux sweet spot for all experiments in this work. A similar mode structure with no flux sensitivity can be achieved, either by breaking the ring by splitting one of the capacitor pads to create two halves with large mutual capacitance, or by using a linear chain of three junctions~\cite{PhysRevA.98.052318}.

Coupling multiple trimons is necessary in order to use them in a scalable architecture. Because trimons have similar charge matrix elements to transmons, they can be directly coupled capacitively, leading to charge-charge interactions between coupled modes. These appear as $YY$ interactions in the qubit space. In such an architecture, inter-trimon $ZZ$ interactions of order $\sim$ MHz would appear. These interactions would be too weak to cause independently-resolvable transition frequencies (unlike the intra-trimon $ZZ$ interactions that our gate schemes rely on). This inter-trimon $ZZ$ interaction would thus be a source of error that must be accounted for just as it is in multi-transmon architectures. One way to suppress this inter-trimon $ZZ$ interaction would be to couple trimons both directly and via tunable transmon couplers similar to those used in standard transmon-based architectures ~\cite{kandalaDemonstrationHighFidelityCnot2021}. Such trimon-transmon capacitive coupling was previously demonstrated in a 3D geometry~\cite{PhysRevApplied.7.054025}. Carefully designing the coupler circuit symmetry could enable it to selectively couple to one, two, or all three modes in either trimon. 
Given the many cross-Kerr interactions that might couple the two trimons, 
it will likely be necessary
to detune the systems and use parametrically driven couplings to avoid frequency crowding~\cite{bertetParametricCouplingSuperconducting2006,reagorDemonstrationUniversalParametric2018,subramanianEfficientTwoqutritGates2023}. Another potential avenue would be to use a double transmon coupler~\cite{DTC_goto, DTC_prx_li, DTC_campbell}, which can eliminate residual coupling between two modes even when they are highly detuned, with the decoupling condition depending only weakly on their individual frequencies. This could then be used in a gate scheme such as an adiabatic CZ gate \cite{chen_unlocking_2026}, with intra-trimon DD used to suppress any residual unwanted interactions when the coupling is on. Developing such multi-trimon gate schemes will be the subject of future work. 

Finally, the trimon qubit modes are likely to experience correlated decoherence due to shared circuit elements, such as common-mode frequency fluctuations ($Z$ noise) affecting all modes simultaneously. For instance, in this device the overlapping dispersive shifts mean that photon number fluctuations in a resonator cause the same frequency shift to all states with the same number of excitations.
This could lead to correlated errors which 
are usually 
seen as 
a challenge 
for quantum error correction. However, 
to the extent that frequency fluctuations are predominantly common-mode, each excitation manifold acquires only a global phase and is therefore (approximately) insensitive to this noise source. This collective error mechanism naturally 
enables efficient error suppression by encoding in a decoherence-free subspace insensitive to common-mode noise~\cite{Zanardi:97c,lidarDecoherenceFreeSubspacesQuantum1998,Kwiat:00,PhysRevLett.91.217904}. Differential frequency fluctuations could be suppressed with DD within this subspace \cite{wuCreatingDecoherenceFreeSubspaces2002,quirozDynamicallyGeneratedDecoherencefree2024}.
Moreover, the dominant 
remaining error channel in such an encoding
is expected to be energy decay, which would take the system out of the subspace. This constitutes an erasure rather than a logical error, opening a clear path toward error correction schemes tailored to erasure-biased noise~\cite{grasslCodesQuantumErasure1997,aliferisFaulttolerantQuantumComputation2008, stephensHighthresholdTopologicalQuantum2013,chaoFlagFaultTolerantError2020,guOptimizingQuantumError2024,chouSuperconductingDualrailCavity2024}. Bit-flip errors within this space have no dipole matrix element and so would likely be rare, requiring higher-order processes. Together, these features position the trimon as a promising platform for realizing intrinsically error-suppressed qubits or qutrits with strongly biased erasure errors, a direction that will be the focus of future work. Likewise, a full exploration of the different error channels of the trimon (e.g., the effects of quasiparticle tunneling) will be the subject of future work.

\subsection*{Acknowledgments}
The authors would like to thank Kishor Salunkhe, Jay Deshmukh, and Arian Vezvaee for useful discussions. VM acknowledges funding from Research Corporation for Science Advancement under Grant SP-BRG-2025-016 and the National Science Foundation, the Quantum Leap Big Idea under Grant No. OMA-1936388. DK, DAL, and EMLF acknowledge funding from the Office of Naval Research under Grant N0001-4-26-12092. SAS acknowledges funding from the Army Research Office under LPS Qubit Collaboratory Grant W911NF-25-1-0255. KS and DAL acknowledge funding from the U. S. Army Research Laboratory and the U. S. Army Research Office under contract/grant number W911NF2310255. RV acknowledges support from the Department of Atomic Energy of Government of India 12-R\&D-TFR-5.10-0100. Devices were fabricated and provided by the Superconducting Qubits at Lincoln Laboratory (SQUILL) Foundry at MIT Lincoln Laboratory, with funding from the Laboratory for Physical Sciences (LPS) Qubit Collaboratory.

\subsection*{Data Availability}
The data supporting the findings of this work are openly available at \cite{FK2/CKCVWQ_2026}. 

\appendix
\renewcommand{\thefigure}{A\arabic{figure}}
\renewcommand{\thetable}{A\Roman{table}}
\renewcommand{\theHfigure}{A\arabic{figure}}
\setcounter{figure}{0} 
\section{Device Theory}
\label{device_theory}
For a generalized four-node multimodal~\cite{PhysRevA.98.052318} circuit with linearized Josephson energy terms, the Lagrangian can be written in matrix form as
\begin{equation}
    \mathcal{L}=\frac{1}{2}\sum_{i,j=1}^{4}\frac{d\Phi_{i}}{dt} C_{ij} \frac{d\Phi_j}{dt}
    -\frac{1}{2\phi_0^2}\sum_{i,j=1}^{4}\Phi_i E_{Lij}\Phi_j,
\end{equation}
where $C$, given by

\begin{equation}
\begin{aligned}
C &= 
\begin{pmatrix}
\sum_i C_{1i} & -C_{12} & -C_{13} &  -C_{14} \\
-C_{21} & \sum_i C_{2i} & -C_{23} & -C_{24} \\
-C_{31} & -C_{32} & \sum_i C_{3i} & -C_{34} \\
-C_{41} & -C_{42} & -C_{43} & \sum_i C_{4i}
\end{pmatrix},\\
\end{aligned}
\end{equation}
is the capacitance matrix, and $E_L$, given by
\begin{equation}
\begin{aligned}
E_L &=
\begin{pmatrix}
\sum_i E_{L1i} & -E_{L12} & -E_{L13} & -E_{L14} \\
-E_{L21} & \sum_i E_{L2i} & -E_{L23} & -E_{L24} \\
-E_{L31} & -E_{L32} & \sum_i E_{L3i} & -E_{L34} \\
-E_{L41} & -E_{L42} & -E_{L43} & \sum_i E_{L4i}
\end{pmatrix}
\end{aligned}
\end{equation}
is the inductive-energy (Laplacian) matrix. 

The capacitance matrix defined above is always positive definite due to the strictly positive ground capacitances $C_{ii}$. Combined with the real-symmetric nature of the inductance matrix $E_L$, we can simultaneously diagonalize both matrices, thereby obtaining the system’s normal modes. 
Specifically, we use the following result: For $C$ and $E_L$ two real symmetric matrices with $C$ positive definite, there exists a non-singular matrix $M$ such that $M^TCM=I$, and $M^T\left(E_L/\phi_0^2\right)M=\Omega$, where $\Omega$ is a real diagonal matrix. Choosing $M=C^{-1/2}A$ for some orthogonal matrix $A$ automatically satisfies $M^TCM=I$. Substituting the same $M$ into the second equation gives $A^T\Psi A=\Omega$, where $\Psi\equiv (C^{-1/2})^T(E_L/\phi_0^2)C^{-1/2}$. Thus, 
$A$ is the orthogonal matrix that diagonalizes $\Psi$, and the second equation 
yields the diagonal matrix $\Omega$, whose diagonal elements are simply the eigenvalues of $\Psi$. Denoting these eigenvalues as $\omega_\mu^2$, we can write the Lagrangian of the system in terms of its normal modes as:
\begin{equation}
\tilde{\mathcal{L}}=\frac{1}{2} \sum_{\mu=0}^{3}\left[\left(\frac{d \tilde{\Phi}_\mu}{d t}\right)^2-\omega_\mu^2 \tilde{\Phi}_\mu^2\right]
\end{equation}
For a generalized four-node multimodal 
circuit there will be 
four normal modes in total. In our case, however, one frequency always vanishes. The matrix $E_L$ is a graph Laplacian (the potential depends only on flux differences), so its rows are linearly dependent and $\det E_L = 0$. From $M^{T}\left(E_L/\phi_0^2\right)M=\Omega$ and the invertibility of $M$ it follows that $\det\Omega = 0$, hence at least one eigenvalue, and thus one frequency, is zero. This zero mode corresponds to a global (common-mode) shift of all node fluxes (a gauge degree of freedom) and can be removed by fixing a reference node; the four-node trimon therefore has three dynamical modes with nonzero frequency.

Extending this linear potential model to the weakly anharmonic potential of a Josephson junction, we obtain
\begin{align}
\begin{split}
U_{\mathrm{pot}} &=-\sum_{i>j=1}^4\left[E_{J i j} \cos \left(\frac{\Phi_i-\Phi_j}{\varphi_0}\right)\right]\\
&\approx-\sum_{i>j=1}^4 E_{J i j} \left[ 1-\frac{1}{2}\left(\frac{\Phi_i-\Phi_j}{\varphi_0}\right)^2 \right.\\
&\left. \hphantom{\approx-\sum_{i>j=1}^4 E_{J i j} \Big[} +\frac{1}{24}\left(\frac{\Phi_i-\Phi_j}{\varphi_0}\right)^4\right]
\end{split}
\end{align}
Consider a nearly symmetric planar trimon with identical Josephson energies $E_J$ and diagonally symmetric capacitor pads. Neighboring node pairs are coupled through capacitances $C_C$, while the diagonal pads have capacitances $C_{13}=C_A$ and $C_{24}=C_B$ between them. In the regime relevant to our device, the ground capacitances satisfy $C_{11} \approx C_{33}$ and $C_{22} \approx C_{44}$, but with $C_{11} \not\approx C_{22}$. Under these conditions, the charging energies of the three normal modes can be obtained analytically, and the corresponding charging-energy prefactors are $E_{C_\mu}=e^2/(2C_\mu^{\mathrm{eff}})$:
\begin{align}
\begin{split}
&E_{C_A}=
\frac{e^2}{2(C_C+C_A)+C_{11}} ,
   \quad
E_{C_B}=
\frac{e^2}{2(C_C+C_B)+C_{22}} ,
\\&
E_{C_C}=
\frac{e^2}{4 C_C+C_{11}+C_{22}+\sqrt{16(C_C)^2+\left(C_{11}-C_{22}\right)^2}}
\end{split}
\label{ecvals}
\end{align}
The effective charging energies are strongly affected by the large pad to ground capacitances $C_{ii}$ (in contrast to a trimon in a 3D cavity, where the ground capacitances are much smaller). The remainder of the theoretical treatment follows the framework of~\cite{tanay_roy}. After canonical quantization of the Lagrangian, at zero external flux $\Phi_{\mathrm{ext}}$, the Hamiltonian, in terms of number operator $n_\mu$, takes the form:
\beq
\mathcal{H}=\sum_{\mu=A,B,C}\left(\omega_\mu\hat{n}_\mu-J_\mu \hat{n}_\mu^2\right)-2\sum_{\mu>\nu}J_{\mu\nu}\hat{n}_\mu \hat{n}_\nu
\label{Hamiltonian_supp}
\eeq
where, 
\begin{equation}
\begin{gathered}
\omega_A = \sqrt{8E_J E_{C_A}}-\beta_A, \qquad
\omega_B = \sqrt{8E_J E_{C_B}}-\beta_B, \\[8pt]
\omega_C = \sqrt{32E_J E_{C_C}}-\beta_C,
\end{gathered}
\end{equation}
\begin{equation}
\beta_\mu=J_\mu+J_{\mu\nu}+J_{\mu\eta,} 
   \quad \text{where}\quad \mu\neq\nu\neq\eta
\end{equation}
\begin{equation}
J_A=\frac{E_{C_A}}{8}\text{,}
   \quad
J_B=\frac{E_{C_B}}{8}\text{,}
\quad
J_C=\frac{E_{C_C}}{2}\text{,}
\end{equation}
and
\begin{equation}
\begin{gathered}
J_{AB}=\frac{\sqrt{E_{C_A}E_{C_B}}}{4}\text{,}
\qquad
J_{BC}=\frac{\sqrt{E_{C_C}E_{C_B}}}{2}\text{,}
\\[6pt]
J_{CA}=\frac{\sqrt{E_{C_A}E_{C_C}}}{2}\text{.}
\end{gathered}
\end{equation}
 The coefficients $J_\mu$ represent the self-Kerr non-linearities ($\Phi_\mu^4$) which add the required anharmonicity to each mode, whereas the cross-Kerr terms $J_{\mu\nu}$ ($\Phi_\mu^2\Phi_\nu^2$) generate the inter-qubit longitudinal coupling that lifts the degeneracy of intermediate levels. 
The transition energy of each qubit, therefore, depends on the occupation number of the other two qubits. This gives us state-dependent transition energies:

\beq
\omega_{A\langle n_B\rangle\langle n_C\rangle}
=\omega_A-\beta_A+(-1)^{n_B}J_{AB}+(-1)^{n_C}J_{CA}
\eeq
\beq
\omega_{\langle n_A\rangle B\langle n_C\rangle}
=\omega_B-\beta_B+(-1)^{n_A}J_{AB}+(-1)^{n_C}J_{BC}
\eeq
\beq
\omega_{\langle n_A\rangle \langle n_B\rangle C}
=\omega_C-\beta_C+(-1)^{n_A}J_{CA}+(-1)^{n_B}J_{BC}
\eeq
\section{Device Design}
\label{app:design}
Circuit parameters were optimized by first approximately solving the circuit Hamiltonian analytically, assuming two-fold mirror symmetry of the circuit and weak nonlinearity. Then the Josephson energy and all capacitance values were swept to find a Hamiltonian where each mode $\mu\in{A,B,C}$ is deep in the transmon limit (the effective Josephson energy of a mode $E_{J_\mu}$ is much larger than the effective charging energy $E_{C_\mu}$); where the anharmonicity is at least 100 MHz for each mode; 
where leakage transitions to higher excited states of each mode are detuned by at least 10 MHz from the set of transitions driven in our experiments (to avoid near-collisions);
where inter-mode dispersive 
(cross-Kerr) conditional transition-frequency splittings
are at least 200 MHz; where the overall mode frequencies lie between 4 and 6 GHz; and where the mode-readout dispersive shifts are all $\sim$ 1 MHz. 
Once circuit parameters are chosen, we perform finite element electromagnetic (FEM) simulations. We use the Ansys Q3D solver to extract the capacitance matrix of the trimon, ground plane, and readout resonator coupling capacitors, and use the Ansys HFSS solver to separately extract the frequency and linewidth of each readout resonator. These simulations follow the method described in Ref.~\cite{shantoSQuADDSValidatedDesign2024}. The simulation outputs and device layout are posted in the SQuADDS database~\cite{shantoSQuADDSValidatedDesign2024,SQuADDSSQuADDS_DBDatasets2024}.

\begin{figure}[ht!]
\centering
\includegraphics[width=\columnwidth]{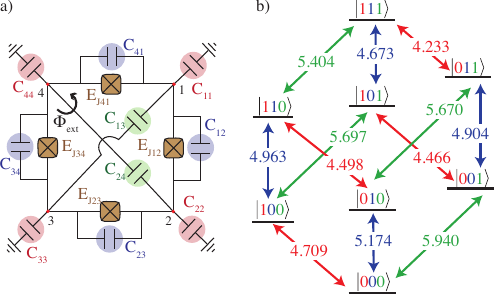}
\caption{(a) Lumped model of the trimon circuit. Red capacitors are shunted to ground. Purple capacitors provide cross capacitance between adjacent nodes, whereas green capacitors, between diagonal nodes, lift the degeneracy of the system. Brown elements are Josephson junctions, which are fabricated to be nominally identical. (b) Measured transition energies of the device used in the experiment. Red, blue, and green arrows correspond to qubit modes A, B, and C, respectively.} 
\label{fig:trimon_circuit}
\end{figure}

\Cref{fig:trimon_circuit} presents the lumped-element circuit model used in our analysis. Electromagnetic simulations of the design layout yield the following capacitance matrix elements:: $C_{11}=46$ fF, $C_{12}=21$ fF,$C_{13}=4$ fF,$C_{22}=30$ fF, $C_{23}=21$ fF, $C_{24}=3$ fF, $C_{33}=53$ fF, $C_{34}=21$ fF, $C_{44}=36$ fF, $C_{14}=21$ fF. 
Using the procedure described in Appendix \ref{device_theory}, we diagonalized the capacitance and inductance matrices constructed from the above values, assuming 
identical junction Josephson energies of $E_J/h = 8.16$~GHz (equivalently a small-signal Josephson inductance $L_J=\phi_0^2/E_J = 20$ nH).
This yields the following mode frequencies: $\omega_A/2\pi=$4.691 GHz, $\omega_B/2\pi=$5.195 GHz, and $\omega_C/2\pi=$5.957 GHz, all within $\pm4$\% of the measured values. The measured device parameters are summarized in \cref{table: device-parameters}.
Another version of the device exhibited larger self- and cross-Kerr couplings, but the present design was selected because it provides more favorable coupling between the trimon modes and the readout resonators. 
As noted above, as long as each mode remains in the transmon regime (effective $E_J/E_C \gg 1$), one can increase $E_C$ (i.e., reduce the effective capacitance) to obtain larger anharmonicities and generally stronger Kerr and cross-Kerr nonlinearities, enabling faster driven transitions while maintaining good spectral isolation.
\begin{table*}[ht!]
\centering
\renewcommand{\arraystretch}{1.3}
\setlength{\tabcolsep}{10pt}

\begin{tabular}{ccccccc}
\toprule
\textbf{Mode} &
${\omega/2\pi}$ (GHz) &
${2J_\mu/2\pi}$ (MHz) &
${2J_{\mu\nu}/2\pi}$ (MHz) &
${T_1}$ ($\mu$s) &
${T_2^{\mathrm{Hahn}}}$ ($\mu$s) &
${\kappa/2\pi}$ (MHz) \\
\midrule
A & 4.709 & 118 & AB, $211$ & 54 $\pm$ 11 & 45 $\pm$ 8 & -  \\
B & 5.174 & 129 & BC, $270$ & 38 $\pm$ 7 & 34 $\pm$ 6 & -  \\
C & 5.940 & 164 & CA, $243$ & 33 $\pm$ 9 & 30 $\pm$ 9 & -  \\
Resonator 1 & 7.667 & - & - & - & - & 1.69 \\ 
Resonator 2 & 8.283 & - & - & - & - & 1.27 \\ 
\bottomrule
\end{tabular}
\caption{Measured parameters of the trimon device used in the experiment.}
\label{table: device-parameters}
\end{table*}

To verify our readout coupling scheme, we performed a circuit-level simulation of the trimon-resonator cross-Kerr shifts in scQubits~\cite{scqubits} using circuit parameters extracted from an Ansys Q3D simulation of the trimon design. The resulting cross-Kerr values agreed well with the experimental data, with less than $10\%$ difference. For the resonator used for readout in this work ($\omega/2\pi=7.667$ GHz), the cross-Kerr values were $\chi_{A00}\approx1.18$ MHz, $\chi_{0B0}\approx1.34$ MHz, and $\chi_{00C}\approx1.36$ MHz.
To improve distinguishability between all modes, this simulation could be extended to an optimization problem in which the cross-Kerr differences between adjacent modes in one and two excitation subspaces is maximized with respect to the coupling capacitance and resonator frequencies for a complete readout. Such optimized devices could then be designed in conjunction with narrowband resonators and read out using multi-tone driving, as demonstrated in transmon-based qudit devices~\cite{qudit_readout_blok}. Alternatively, broadband resonators combined with Purcell filters could be used to better distinguish the states.

In either case, we expect that using two resonators would make it easier to distinguish population blobs via simultaneous readout, with each resonator resolving a different part of the state subspace. In addition to the symmetric coupling used in our device, one could also explore asymmetric coupling schemes, in which the resonator coupling capacitor is coupled not to a single pad but to two adjacent pads, coupling directly to modes A, B, and C simultaneously. Using one resonator with symmetric coupling and a second with asymmetric coupling may allow each resonator to resolve a different part of the state subspace, further improving overall distinguishability. Optimizing three-mode readout in the trimon for different device applications will be the subject of future work.

\section{Readout}
\label{readout-appendix}
The nearly-equal dispersive shifts of the three qubits on each resonator make it difficult to distinguish states with the same number of excitations. 
This was a deliberate design choice for this specific device (where readout was optimized for a different application not discussed in this work, see Appendix \ref{app:design}) and is not intrinsic to the trimon. In our device, four rounds of measurement with mapping pulses can still distinguish all eight states~\cite{PhysRevApplied.7.054025,PhysRevApplied.14.014072}. 

This procedure requires eight unique mapping and readout pulses. 
Our control electronics did not allow more than $10$
distinct frequencies for pulses
to be defined in a single experiment. 
Unfortunately, this 
rules out many 
experiments (e.g., three-qubit tomography requires pulses at $14$ frequencies). Given these constraints, in the present work (except for qudit measurements and single-qubit QPT), we restricted ourselves to two-qubit readout using only one resonator at a time while always keeping the third qubit in the ground state.

\begin{figure*}[ht!]
\centering
\includegraphics[width=\textwidth]{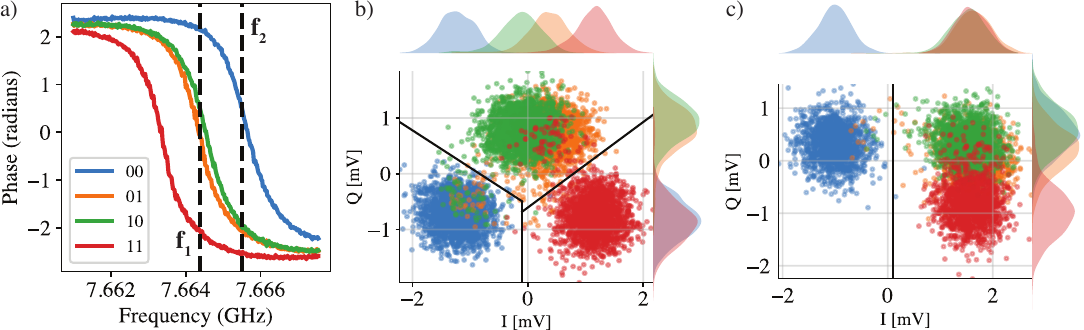}
\caption{Readout scheme used in our setup. (a) Dispersive phase shift of the readout resonator when the qubits occupy different computational states. The readout frequency  $f_1$ is used for state tomography measurements, where two rounds of measurements are required to fully determine the state. The readout frequency $f_2$ is used for process tomography and qudit experiments, where we only need to distinguish whether the final state is $\ket{00}$ or not. (b) IQ distributions of the four computational states measured at frequency $f_1$. Colors correspond to the same states as in panel (a). The black lines indicate the decision boundaries used for state assignment. (c) IQ distributions of the same states measured at frequency $f_2$. The black line indicates the boundary separating the $\ket{00}$ cluster from the others, enabling high-fidelity readout of the ground state.}
\label{fig:readout}
\end{figure*}

Before each experiment, we obtain single-shot measurement histograms by preparing each of the four basis states individually many times and recording their corresponding measurement outcomes. We then apply a Support Vector Classifier (SVC)  algorithm to define demarcation boundaries that separate the different excitation subspaces, $n=0$ ($\ket{00}$), $n=1$ ($\ket{01}$ or $\ket{10}$), and $n=2$ ($\ket{11}$) (\cref{fig:readout}b). These boundaries were subsequently used to assign probabilities for each excitation number $n$ after executing the experiment. 

To read out an arbitrary state, we first prepare the state and measure it using a readout tone at $f_1$ (\cref{fig:readout}a), which resolves the $\ket{00}$ and $\ket{11}$ populations, according to the decision boundaries defined above. Measurement outcomes falling within the $\ket{01}$ or $\ket{10}$ region are discarded. After waiting for an appropriate reset time, the same state is re-prepared. Now, before sending in the readout tone, we apply projection pulses [$X_{\pi}$ at $\omega_{0B0}$ and $\omega_{1B0}$ (or $\omega_{A00}$ and $\omega_{A10}$)] to move the population between states: $\ket{00} \rightarrow \ket{01}$ and $\ket{10} \rightarrow \ket{11}$ (or $\ket{00} \rightarrow \ket{10}$ and $\ket{01} \rightarrow \ket{11}$). This effectively maps the populations of $\ket{01}$ and $\ket{10}$ onto the $\ket{00}$ and $\ket{11}$ measurement blobs, allowing us to infer their populations. By combining the results from these two measurement rounds, we reconstruct the full statistics for a given state. We note that it is possible for this procedure to give some probabilities that are not equal to 1 due to statistical fluctuations, and so we use a very large number of shots (40,000 typically) to ensure statistical fluctuations are negligible. This procedure provides $93\%$ average SPAM fidelity, where the majority of the error is due to measurement, as our gates have high fidelity. 

For experiments in which only the population of the $\ket{000}$ state is required, such as in process tomography or qudit measurements, we perform readout using the tone at frequency $f_2$(\cref{fig:readout}a), where the $\ket{00}$ cluster is well separated from the other basis states, allowing a clear decision boundary to be drawn (\cref{fig:readout}c). This yields a measurement fidelity close to 98\% for the ground state. For qudit measurements, errors are dominated by state preparation of the initial qudit state. We therefore divide off this initial fidelity to define the fidelity as 1 at the beginning of the sequence.

To read out a single qubit while tracing out the other qubits' states, we apply two mapping pulses so that the main qubit's $\ket{0}$ states are mapped to the 1- and 3-excitation subspaces while the qubit's $\ket{1}$ states are mapped to the 0- and 2-excitation subspaces. For instance, to read out qubit B while tracing out qubits A and C, we apply $ccX_\pi^{0B0}$ to map $\ket{000}\leftrightarrow\ket{010}$, creating a 1-excitation space where all states (formerly) had B in the $\ket{0}$ state. Likewise, we apply $ccX_\pi^{1B1}$ to map $\ket{101}\leftrightarrow\ket{111}$, creating a 2-excitation space where all states (formerly) had B in the $\ket{1}$ state while mapping the remaining $\ket{0}$ state to the 3-excitation state. We then simultaneously readout to distinguish the 0-, 1-, and 2-or-3-excitation states with one resonator and the 0-or-1-, 2-, and 3-excitation states using another drive with slightly different frequency applied to the same resonator, with the frequency chosen to optimize SNR. By comparing the populations, we can thus read out the population of qubit B in a single shot while remaining insensitive to the states of qubits A and C.

\begin{figure}[ht!]
\centering
\includegraphics{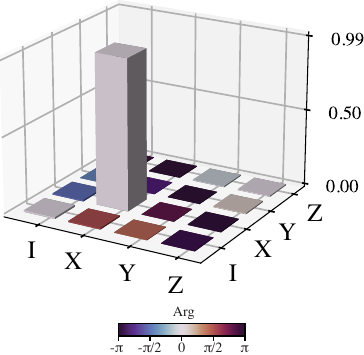}
\caption{Single-qubit QPT for a 60 ns unconditional X gate on qubit B. QPT is performed after the system is initialized in the state $\ket{+}_A \ket{0}_B \ket{+}_C$. The readout process measures the state of qubit B while tracing out the remaining qubits. Bar height indicates the magnitude of each matrix element, and color denotes its phase.}
\label{fig:single_qubit_QPT}
\end{figure}

For all experiments where gate or state fidelity is being measured, we correct for SPAM error. 
Our SPAM mitigation for state fidelities uses calibration data obtained by preparing reference states with CC$\mathcal{R}_\pi$ gates and measuring them; as a result, imperfections in these preparation pulses can be partially absorbed into the inferred SPAM model, introducing some ambiguity in the interpretation of SPAM-corrected state fidelities.
For example, when characterizing the state fidelity of the Bell state prepared using a $ccX_{\pi}^{A00}$ gate followed by a Raman $\sqrt{i\mathrm{SWAP}}$ gate as in \cref{fig:QPT_raman}(c), we correct for the error involved in the first gate. The state fidelity number may thus be interpreted as capturing errors due to the $\sqrt{i\mathrm{SWAP}}$ gate only. Details of the SPAM correction procedure for both QST and QPT are given Appendix \ref{section-tomo}. 


\section{Cable phase and Gate Calibration}
\label{Gate-calibration}
Because pulses for each mode are generated on separate DAC channels and then upconverted before being combined onto a single physical line, they arrive at the trimon with different relative phases. This requires a cable phase calibration to ensure that pulses arriving at the device maintain correct relative phases. The issue can be illustrated by considering only two modes, A and B, with drive phases $\phi_{A00}$, $\phi_{0B0}$, $\phi_{A10}$, and $\phi_{1B0}$ for the transitions $A00$, $0B0$, $A10$, and $1B0$, respectively. We first prepare $\ket{110}$ via two distinct pulse sequences, each traversing a different path on the Bloch sphere, $\ket{000}\rightarrow\ket{100}\rightarrow\ket{110}$ and $\ket{000}\rightarrow\ket{010}\rightarrow\ket{110}$, implemented using a pair of full $X_\pi$ pulses. Since each sequence is separate, the relative phase between the two paths plays no role, and both sequences yield the same final population in $\ket{110}$, as we confirm experimentally. We then repeat this experiment, but decompose each $X_\pi$ pulse into two $X_{\pi/2}$ pulses and interleave the two sequences, so that $\ket{110}$ is reached through a combination of both paths, with pulses from one sequence played in between pulses from the other. In this interleaved sequence, the final state now depends on the accumulated relative phase. If $\phi_{A00}+\phi_{1B0}\neq\phi_{0B0}+\phi_{A10}$, the second $X_{\pi/2}$ at $\omega_{1B0}$ acts along a shifted axis, resulting in incomplete population transfer to $\ket{110}$.

To correct this, we fix mode A as our phase reference and sweep the phase of the pulse driving $\omega_{1B0}$ until the condition $\phi_{A00}+\phi_{1B0}=\phi_{0B0}+\phi_{A10}$ is satisfied. At this condition, we recover complete population transfer to $\ket{110}$, matching the level achieved in the non-interleaved experiment. We repeat this calibration procedure for the remaining transitions, requiring a total of six phase calibration experiments. Mode A's phase is taken as the reference throughout, 
with each experiment yielding a relative phase-matching condition, one of which fixes the offset. These calibration parameters are fixed for a given setup, independent of pulse length or amplitude.

\begin{figure}[ht!]
\centering
\includegraphics[width=\columnwidth]{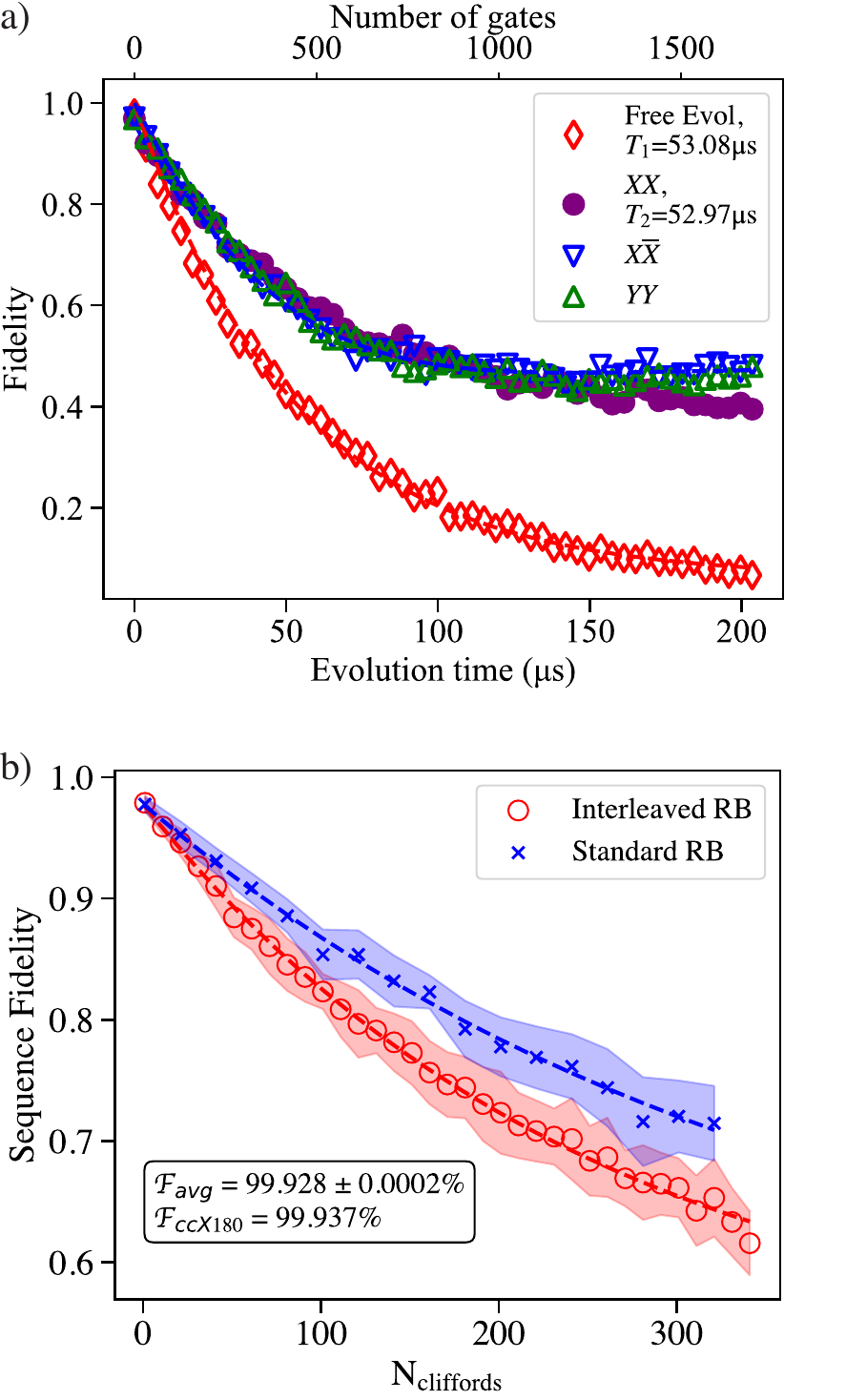}
\caption{(a) Deterministic Benchmarking (DB) of a $X_\pi$ gate at $\omega_{A00}$, implemented using a 60 ns cosine DRAG pulse envelope. (b) Randomized Benchmarking (RB) of a $X_\pi$ gate at $\omega_{0B0}$, implemented using a 40 ns cosine DRAG pulse envelope.}
\label{fig:DB-RB}
\end{figure}

To calibrate gates, we use the deterministic benchmarking (DB) procedure~\cite{DB}. This involves preparing an even superposition of the two states or spaces the gate rotates between, then repeating pairs of two of the same $\pi$ rotations between these states or spaces, or pairs of a $\pi$ rotation and its inverse, to amplify error terms. In particular, we applied the $YY$ and $X\overline{X}$ dynamical decoupling sequences to identify and correct over/under rotation and phase errors, respectively. If the gate is effectively a $\pi/2$ rotation, then we 
apply
four gates or two gates and two inverses. We then measure the fidelity to the initial state. In all cases, the nominal operation is identity. Oscillations in fidelity come from coherent errors in the rotation angle (for pairs of the same rotation) or rotation axis (for rotation/inverse pairs). \Cref{fig:DB-RB}a shows a DB trace for a $X_{\pi}$ gate with 60ns gate time at the $\omega_{A00}$ transition. Pulse parameters were iteratively adjusted until the oscillations in both the $YY$ and $X\overline{X}$ sequences were suppressed, indicating properly calibrated gate amplitudes and phases with minimal coherent errors. The same procedure was used to calibrate all transitions. For calibrating higher-level transitions, the same DB sequences were applied on the states connected by those transitions. \cref{fig:single_qubit_QPT} shows single-qubit QPT for a 60ns unconditional X gate on qubit B. In this case, pulses at $\omega_{0B0}$, $\omega_{0B1}$, $\omega_{1B0}$, and $\omega_{1B1}$ were applied simultaneously and calibrated using the same procedure, yielding a QPT gate fidelity of 99.56\%. 

To benchmark gates against incoherent errors, we performed Randomized Benchmarking (RB) experiments on all $\omega_{00}$ transitions. An RB decay curve for a 40ns $X_{\pi}$ gate at $\omega_{0B0}$ is shown in \cref{fig:DB-RB}b. For the 60ns $X_{\pi}$ gate at $\omega_{A00}$ and $\omega_{00C}$, we obtained average gate fidelities of $99.949\pm 0.02$\% and $99.872\pm0.003$\%, respectively. These fidelities are close to decoherence limited and are expected to improve with better design and fabrication techniques.
\section{Tomography}
\label{section-tomo}
\subsection{Quantum state tomography}
The density matrix of two qubits can be reconstructed by performing 
measurements in the $9$ local Pauli bases, i.e., all combinations $\Sigma_{\mathrm{q_1}}\otimes \Sigma_{\mathrm{q_2}}$ with $\Sigma_{\mathrm{q_1,q_2}}\in\{X,Y,Z\}$~\cite{Kundu2019, DariaNM2024}. 
To correct for SPAM errors, we construct a 
$4\times4$ confusion matrix
by preparing each computational basis state and recording the assigned outcome probabilities using the same pulse scheduling and state-assignment procedure as in the tomography experiment. These outcome-probability vectors form the columns of the confusion matrix. The confusion matrix is constructed before each tomography run. The average of all the diagonal elements of the confusion matrix was approximately $93\%$, with the ground state fidelity being the highest at $94.5\%$.
For an arbitrary input state, each of the $9$ measured probability vectors is multiplied by the inverse of the confusion matrix to obtain the SPAM corrected probabilities. 
With all 
$9$ corrected expectation values, the complete two-qubit density matrix is reconstructed via Maximum Likelihood Estimation (MLE) following these steps: 

\begin{enumerate}

\item Define a lower-triangular matrix, $T$ with complex off-diagonal elements, where the density matrix can be constructed as $\rho = \frac{T^{\dagger} T} {\tr(T^{\dagger} T)}$. This assures the Hermiticity of $\rho$ as well as its trace-normalization.

\item Construct a residual vector of the form $r_{k} = \tr(\hat{\mathcal{M}}_{k}\rho) - p_{k}$, where $\hat{\mathcal{M}}_k$ represents the measurement operators (e.g., the Pauli-product operator basis including $I\otimes I$) and $p_k$ are the corresponding experimentally estimated expectation values.

\item Employ the least\_squares function under the scipy.optimize package \cite{SciPy} to minimize the 
residual vector $\{r_k\}$
to extract the optimized values for the $T$ matrix elements and reconstruct the density matrix.  

\end{enumerate}

\subsection{Quantum process tomography} 

We begin by preparing Bloch sphere states on each qubit using the set of operations $\{I, \mathcal{R}_x(\pi),\mathcal{R}_x(\pm\pi/2),\mathcal{R}_y(\pm\pi/2)\}$. After state preparation, we 
apply the target gate
and subsequently perform basis changes using the same set of rotations. For each configuration, we measure the probability of both qubits being in their ground state. This procedure constitutes an over-complete set of measurements needed for the process tomography and yields an estimate for the process matrix $\hat{\chi}_{G}$ for the gate under consideration. To correct our QPT results for SPAM error, we interleave the QPT experiments with an additional reference experiment in which we do not 
apply the target gate
while all state preparation and basis change operations are kept identical. The additional reference experiment yields $\hat{\chi}_I$, which should ideally correspond to a process matrix for an identity gate, and contains information about the SPAM errors~\cite{EasybetterQPT}. To account for SPAM errors in $\hat{\chi}_{G}$, we calculate an error superoperator using $\hat{\chi}_I$, and use it to modify our ideal SPAM matrices. We distribute the error equally between state preparation and measurement. Then, using these modified SPAM matrices for inversion, we obtain a new SPAM corrected estimate of $\hat{\chi}_{G}$. The QPT gate fidelity is computed using
$\mathcal{F}_G:=\frac{d\mathcal{F}_P(\chi_G,\chi)+1}{d+1}$, where $\chi$ is the target process matrix, $d$ is the Hilbert space dimension, and $\mathcal{F}_P(\chi_G, \chi):=\text{Tr}(\sqrt{\sqrt{\chi}\chi_G\sqrt{\chi}})^2 $ denotes the process fidelity. The average gate fidelity obtained for the reference identity process $\hat{\chi}_I$ was approximately $93\%$. 

\section{Qudit Dynamical Decoupling}
\label{Qudit-DD-appendix}
\begin{figure}[ht!]
\centering
\includegraphics{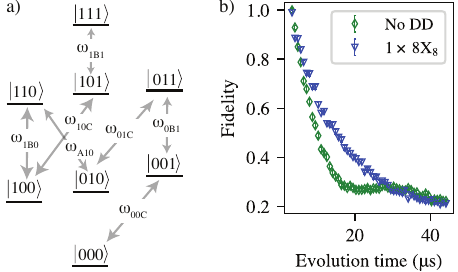}
\caption{(a) Energy level structure and the corresponding transitions for $X_8$ gate compilation used in qud8 dynamical decoupling (DD) experiment. An effective X pulse is composed of $7$ native $ccX$ gates. (b) Experimental results for the qud8 single-axis DD sequence. The two sequences shown were interleaved and executed as a single experiment. The fidelity is normalized by the maximum fidelity achieved in the experiment, defined as the maximum value among the two sequences. Error bars correspond to $\pm2\sigma$.}
\label{fig:qud8_DD}
\end{figure}

Since 
a significant 
error mechanism in transmons and similar circuits is dephasing due to $1/f$ noise~\cite{PRXQuantum.5.010320}, we focus on single-axis dynamical decoupling sequences. If the orthogonal basis states are labeled as $\ket{i}$, the required decoupling sequence simplifies to repeated application of the basic shift operator $X_d = \sum_{k=0}^{d-1} \ketbra{(k+1)\! \mod{d}}{k}$, where $d$ is the dimension of the 
chosen $d$-level subspace; we refer to this as
the $dX_d$ sequence. 
The choice of states for the Hilbert space and the compilation for the $X_d$ gate are given in \cref{table:dd}. Experimental results for the
qud8 single-axis DD sequence is given in \Cref{fig:qud8_DD}. 
\begin{table*}[ht!]
\setlength{\tabcolsep}{12pt}
\centering
\begin{tabular}{c c c}
\toprule
$d$ & Ordered basis for Hilbert space & $X_d$ compilation \\
\midrule
3 & $\{\ket{000}, \ket{100}, \ket{010}\}$ & $CCX_{0B0} \circ CCX_{A00}$ \\
4 & $\{\ket{000}, \ket{100}, \ket{110}, \ket{010}\}$ & $CCX_{A00} \circ  CCX_{1B0} \circ CCX_{A10}$ \\
6 & $\{\ket{100}, \ket{000}, \ket{001}, \ket{011}, \ket{010}, \ket{110}\}$ & $CCX_{A00} \circ  CCX_{00C} \circ  CCX_{0B1} \circ  CCX_{01C}\circ CCX_{A10}$ \\
8 & $\{\ket{000}, \ket{001}, \ket{011}, \ket{010}, \ket{110}, \ket{100}, \ket{101}, \ket{111}\}$ & $CCX_{00C} \circ  CCX_{0B1} \circ  CCX_{01C} \circ  CCX_{A10}\circ CCX_{1B0}$\\ & & $\circ CCX_{10C}\circ CCX_{1B1}$ \\
\bottomrule
\end{tabular}
\caption{Choice of Hilbert space and $X_d$ gate compilation for different $d$ = dimension of Hilbert space values. $CCX_\Gamma$ denotes the $CCX180$ gate between the energy levels connected by transition $\Gamma$.}
\label{table:dd}
\end{table*}

\section{Measurement setup}
\begin{figure*}[ht!]
\centering
\includegraphics[width=\textwidth]{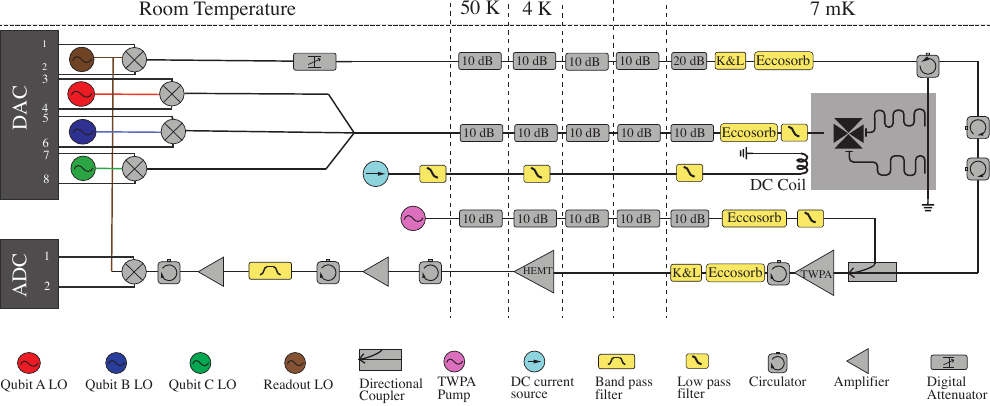}
\caption{Room-temperature measurement and cryogenic setup of the experiment.}
\label{fig: measurement setup}
\end{figure*}

\Cref{fig: measurement setup} shows the room temperature and cryogenic setup of the experiment. The device is packaged in a copper box and surrounded by an additional copper can as well as a Cryoperm shielding to protect the device from infrared radiation and external magnetic fields. The device is further thermalized to the mixing chamber stage via a copper plate. The coaxial lines are thermalized via cryogenic microwave attenuators. Note that the very last attenuators are specific cryogenic attenuators (QMC-CRYOATTF) known for their thermalization properties to minimize the impact of thermal noise injected into the system.
The drive line has $55$ dB of fixed attenuation to attenuate the ambient noise entering the fridge, in addition to roughly 10 dB of distributed attenuation in the coaxial cables. This arrangement of the attenuators allows us to achieve Rabi oscillations at a desirable rate for our experiments while minimizing noise. In addition, we installed $8.4$ GHz low-pass filters (MiniCircuits VLFX $8400$+) to mitigate high-frequency noise. Finally, for the readout input line, we added $60$ dB of attenuation with a KNL low-pass filter (LPF) at $12$ GHz. An Eccosorb infrared filter (QMC-CRYOIRF-002MF) was installed for every single microwave line inside the copper shielding with $> 10$ GHz cutoff frequencies to absorb the infrared radiation. In order to flux tune the device, we make use of a superconducting coil biased via DC twisted pairs in the fridge. In addition to the RC filter at room temperature, the DC lines also run through both RC and LC filters at 4K and milliKelvin stages, respectively, to attenuate the high frequency noise picked by the voltage source. The arbitrary waveforms used to control the experiments in this work were generated using a Quantum Machine OPX1. The waveforms were then upconverted using conventional IQ mixers to control the qubits at higher frequencies.

To amplify the output signal, we use a high-electron-mobility transistor (HEMT) low-noise amplifier at the 4K stage as well as a traveling-wave parametric amplifier (TWPA) at milliKelvin temperatures with gains of about $40$ dB and $30$ dB, respectively. The output signal is then filtered using a band pass filter to avoid saturating the room temperature amplifiers due to the presence of the TWPA pump signal. The amplified signal is finally downconverted and digitized using a heterodyne scheme to readout the state of the qubit.
\FloatBarrier

\bibliography{references}
\end{document}